\documentstyle[12pt,psfig]{article}

\def\al{\alpha}
\def\be{\beta}
\def\ga{\gamma}

\def\ep{\epsilon}

\def\et{\eta}

\def\ka{\kappa}

\def\rh{\rho}

\def\ph{\phi}

\def\om{\omega}
\def\Ga{\Gamma}
\def\De{\Delta}

\def\ket#1{|{#1}\rangle}

\def\frac#1#2{{\textstyle{{#1}\over {#2}}}}
\def\lsim{\mathrel{\rlap{\lower4pt\hbox{\hskip1pt$\sim$}}
    \raise1pt\hbox{$<$}}}
\def\gsim{\mathrel{\rlap{\lower4pt\hbox{\hskip1pt$\sim$}}
    \raise1pt\hbox{$>$}}}
\def\sqr#1#2{{\vcenter{\vbox{\hrule height.#2pt
         \hbox{\vrule width.#2pt height#1pt \kern#1pt
         \vrule width.#2pt}
         \hrule height.#2pt}}}}
\def\Re{\hbox{Re}\,}
\def\Im{\hbox{Im}\,} 

\newcommand{\beq}{\begin{equation}}
\newcommand{\eeq}{\end{equation}}
\newcommand{\bea}{\begin{eqnarray}}
\newcommand{\eea}{\end{eqnarray}}
\newcommand{\rf}[1]{(\ref{#1})}
 
\renewenvironment{thebibliography}[1]
 { \rm
   \begin{list}{\arabic{enumi}.}
    {\usecounter{enumi} \setlength{\parsep}{0pt}
     \setlength{\itemsep}{3pt} \settowidth{\labelwidth}{#1.}
     \sloppy
    }}{\end{list}}
 
\begin{document}
\titlepage
 
\baselineskip=20pt

\begin{flushright}
{IUHET 437\\}
{IUNTC 01-03\\}
{JLAB-THY-01-20\\}
{June 2001\\}
\end{flushright}

\vglue 1cm

\begin{center}
{{\bf BACKGROUND ENHANCEMENT OF CPT REACH\\
AT AN ASYMMETRIC $\ph$ FACTORY\\}
\vglue 1.0cm
{Nathan Isgur,$^a$ V.\ Alan Kosteleck\'y,$^b$
and Adam P.\ Szczepaniak$^{b,c}$\\} 

\bigskip
{\it $^a$MS 12H2, Jefferson Lab\\}
{\it Newport News, VA 23606, U.S.A.\\}

\bigskip
{\it $^b$Physics Department, Indiana University\\}
{\it Bloomington, IN 47405, U.S.A.\\}

\bigskip
{\it $^c$Nuclear Theory Center, Indiana University\\}
{\it Bloomington, IN 47405, U.S.A.\\}
}

\vglue 0.8cm
  
\end{center}

\bigskip

{\rightskip=2pc\leftskip=2pc\noindent
Photoproduction of neutral-kaon pairs 
is studied from the perspective of CP and CPT studies. 
Interference of the $P$ and $S$ waves,
with the former due to diffractive $\phi$ production
and the latter to $f_0$/$a_0$ production,
is shown to enhance the CPT reach. 
Results are presented of Monte Carlo studies 
based on rates expected in future experiments. 
}

\vskip 1 cm

\baselineskip=20pt
\newpage

\noindent 
{\it 1.\ Introduction.}
Neutral-meson oscillations provide a sensitive tool 
for testing CPT symmetry 
\cite{lw}.
Impressive bounds have been achieved in experiments
with both $K$ mesons
\cite{kexpt,k99}
and $B_d$ mesons
\cite{bexpt}.
Although the CPT theorem
\cite{sachs}
guarantees that the standard model preserves CPT 
by virtue of its construction as a 
Lorentz-invariant quantum field theory,
violations of CPT 
could be exhibited in a more fundamental description
of nature that incorporates physics at the Planck scale
\cite{cpt98}.
For example,
CPT and Lorentz violation may naturally  
arise in string theory
\cite{kps}
and could lead to effects in the various neutral-meson systems
\cite{kp}.
It is therefore valuable to identify 
additional possibilities for future experiments 
with exceptional sensitivity to CPT violation.

In the $K$ system,
CPT bounds of a few parts in $10^{19}$ now exist
on the ratio of the kaon-antikaon mass difference
to the kaon mass 
\cite{kexpt,k99}.
Future improvements over these bounds 
are expected from specialized experiments at kaon factories,
such as KLOE at DAPHNE
\cite{kloe}.
Indeed,
it has long been recognized that 
high-luminosity $\ph$ factories 
are ideal for studies of CP and CPT violation
\cite{daf}
because a decaying $\phi$ meson produces a well-defined flux 
of $C$-odd $K_S K_L$ pairs.

In the present work,
we study the possibility 
that interference effects between the $C$-odd wave 
and a coherent $C$-even $K^0\bar K^0$ background
could be used to enhance further the sensitivity
to parameters describing the weak $K^0$ and $\bar K^0$ decay amplitudes. 
At an $e^+e^-$ collider,
the $\ph$ mesons are essentially the only source 
of $K^0\bar K^0$ pairs. 
However,
such studies potentially suffer from a $C$-even background from the decay 
\cite{gamma}
$\phi\to \gamma f_0(980) \to \gamma K^0\bar K^0$.
Fortunately, 
recent measurements by the SND and CMD-2 collaborations report 
\cite{ra}
a branching ratio
B$(\phi\to\gamma f_0) = 1.9-3.5 \times 10^{-4}$, 
which is insignificant for CP and CPT studies 
\cite{daf2}.  

Here,
we consider instead hadronic production,
where a different mechanism exists.
In particular, 
in photoproduction near the $\ph$ meson peak,
a significant $S$-wave background has been measured 
and attributed to the decays of the $f_0(980)$ and $a_0(980)$ mesons
\cite{fr,bar}.
In what follows,
we show how differences in the angular distributions 
of the $S$ and $P$ waves could be exploited for CP and CPT studies. 

Nature is believed to be described at low energies by
a quantum field theory.
The CPT theorem then suggests that any violations of CPT invariance 
must be accompanied by Lorentz violation.
At the level of the standard model,
small violations of Lorentz and CPT invariance can be introduced
via additional terms in the lagrangian,
which yields a general standard-model extension
\cite{ck}.
In this general context,
it turns out that
neutral-meson oscillations provide a unique sensitivity
to a class of parameters for CPT violation
associated with flavor-changing effects in the quark sector
\cite{ak}.
In fact,
to leading order in small parameters,
tests with neutral mesons are independent
of all others performed to date,
including comparative measurements 
in Penning traps \cite{penning},
spectroscopy of hydrogen and antihydrogen \cite{hhbar},
measurements of muon properties
\cite{muon},
clock-comparison experiments
\cite{ccexpt},
observations of the behavior of a spin-polarized torsion pendulum
\cite{spinpol},
measurements of cosmological birefringence \cite{photon,ck},
and observations of the baryon asymmetry \cite{bckp}.

In the $K$ system,
experiments have reached a sensitivity
of parts in $10^{20}$ to certain parameters for CPT violation
in the standard-model extension \cite{kexpt,k99,ak}.
At leading order,
four coefficients control CPT violation for kaons.
Experiments can therefore place four independent CPT bounds,
only two of which have been obtained to date.
The photoproduction of $\ph$ mesons offers a distinct 
experimental arena with the potential to bound
new combinations of parameters,
a possibility which may well merit careful investigation.
However,
with the exception of a few remarks below,
the scope of the present exploratory work is limited to the
demonstration that a background-enhanced CPT reach is possible
in photoproduction.

\vglue 0.4cm
\noindent 
{\it 2.\ Theory for $S$-$P$ interference.}
In the reaction  
$\ga p \to X p \to K^0 \bar K^0 p$ 
with $J^{PC}(X)=1^{--}(\phi)$, $0^{++}(f_0)$ or $0^{++}(a_0)$,
the $K^0\bar K^0$ wave function 
in the rest frame of the pair can be written as  
\beq
\ket{i} 
= h \int {d^2\hat q}\left[a^P({\hat q})
\ket{ K_S({\hat q})K_L(-{\hat q})}
+ a^S({\hat q})\left( \ket{K_L({\hat q})K_L(-{\hat q})}
-\ket{K_S({\hat q})K_S(-{\hat q})} \right)
\right].
\label{i}
\eeq
Here, 
the first term represents the $P$ wave
and comes from the $\phi$ decay 
and other coherent odd-parity backgrounds
with $a^P(\hat q) = \sum_m a^1_{m} Y_{1m}(\hat q)$,
where $\hat q$ is a unit vector.
The second term comes from $S$-wave 
and other even-parity backgrounds 
with $a^S = \frac{1}{2}a^0_{0} Y_{00}(\hat q)$.
The photoproduction amplitudes 
$a^J_{m}=a^J_{m}(s,t,m_{K\bar K})$
describe the dynamics of the production process.
Since $K\bar K$ photoproduction 
is dominated by helicity-nonflip diffraction
for which the $a^1_{\pm1}$ coefficients are the largest, 
the subsequent $K\bar K$ evolution is best studied 
in the $s$-channel helicity system,
with the $z$-axis defined in the direction 
opposite to the direction of flight of the outgoing
nucleon in the $K\bar K$ rest frame 
\cite{sch}. 
In general an incoherent $S$-wave background is also present
but is largely irrelevant for our analysis.
We comment on its effect later.

The normalization $h$ in Eq.\ \rf{i}
is chosen so that the $K\bar K$ photoproduction rate 
is given by
\beq
{ {dN_{K\bar K}}\over {dt dm_{K\bar K} d^2\hat q}}   
= {\em F} {{d\sigma(\gamma p \to K\bar K p) }\over {dt
dm_{K\bar K}}} W(\hat q),
\eeq
where ${\em F}$ is the photon flux. 
In this equation,
the $K\bar K$ angular distribution $W(\hat q)$ is
taken to be unit normalized,
$\int d{\hat q} W(\hat q) = 1$,
and is given by
\beq
W(\hat q) = \sum_{JJ'}W_{JJ'} = 
\sum_{Jm,J'm'} \rho^{JJ'}_{mm'}  Y_{Jm}Y^*_{J'm'}. 
\label{W}
\eeq
Here,
$\rho^{JJ'}_{mm'} \equiv a^J_{m} a^{J'*}_{m'}$ 
are elements of the spin density matrix.
Summation over nucleon helicities is implicit. 

The two kaons decay into final states $f_\al$,
$\al=1,2$,
with amplitudes given by 
\beq
A(K_{S(L)}\to f_\alpha) 
= a_{\alpha,S(L)} \exp(i\phi_{\alpha,S(L)})
exp\big(-i m_{S(L)}t_\alpha - 
\Ga^\al_{S(L)}t_\al/2\big),
\label{weak}
\eeq
where $\Gamma^\alpha_{S(L)}$ are 
the corresponding partial decay widths. 
For convenience, 
we define as usual for each fixed $\al$ the parameter
\beq
\et_\al \equiv |\et_\al | \exp(i \ph_\al) = a_{\al,L}/a_{\al,S} 
\label{b}
\eeq
as the ratio of the amplitude for the transition
between $f_\al$ and $K_L$ 
to that between $f_\al$ and $K_S$.
For the moment, 
the $f_\al$ are kept general.

The production rate $R(t_1,t_2;\hat q)$
of the final state $f_\alpha$
with momentum direction of $f_1$ 
specified by the solid angle $\hat q$ is 
\beq
R(t_1,t_2,\hat q) 
 = \int dm_{K\bar K} dt { {dN_{K{\bar K}}(f_1,f_2)} \over {dt dm_{K\bar
 K} d^2\hat q}}.
\eeq
This expression can be expanded as 
\beq
R(t_1,t_2,\hat q) = N_{K\bar K} \big( |M_{PP}|W_{PP} +
\sum_{m=0,1} |M^m_{PS}|
W^m_{PS} + |M_{SS}|W_{SS}\big),
\label{ang}
\eeq
where $N_{K\bar K}=F \sigma(\gamma p\to K\bar K p)$ 
is the rate for kaon pair production
integrated over $t$ and $m_{KK}$ in the region of the $\phi$ peak. 
The two angular distributions
$W_{PP}$ and $W_{SS}$ are determined by Eq.\ \rf{W}.
The remaining ones are 
\beq
W^0_{PS} = {\sqrt{3}\over {4\pi}}|\rho^{10}_{00}|\cos\theta,\;\;
W^1_{PS} = -{\sqrt{6}\over
  {4\pi}}|\rho^{10}_{10}|\sin\theta\cos\phi, 
\eeq
assuming that $P$ and $S$ waves are produced 
via natural-parity $t$-channel exchanges,
i.e., 
via Pomeron and $\rh$ or $\om$ mesons,
respectively 
\cite{as}. 
In the general case there is an additional contribution to
$S$-$P$ interference, 
identical in form to the second term in Eq.\ \rf{ang} 
but with $\cos\phi$ replaced by $\sin\phi$ in $W^1_{PS}$ 
and with a shifted phase $\phi^1_B \to \ph^1_B + \pi/2$
(see below). 
The coefficients $M_{JJ'}$ are given by 
\bea
\hbox{\hskip -30pt}
|M_{PP}|&=&\Gamma^1_S\Gamma^2_S e^{-{{\Gamma T}\over 2}} \left[
|\et_2 |^2 e^{\Delta\Gamma {t\over 2}}
+|\et_1 |^2 e^{-\Delta\Gamma {t\over 2}}
-2|\et_1\et_2 |{\Re} e^{i (\De m t +  \De \ph)}
\right],
\nonumber\\
\hbox{\hskip -30pt}
|M_{SS}| &=& \Gamma^1_S\Gamma^2_S\left[ e^{-\Gamma_S T} 
+|\eta_1\eta_2|^2e^{-\Gamma_L T} 
 - 2 |\eta_1\eta_2|e^{-\Gamma {T\over 2}} {\Re} e^{i(\Delta m T
   -\phi_1-\phi_2)}\right],
\nonumber\\
\hbox{\hskip -30pt}
|M^m_{PS}| &=& 
2\Gamma^1_S\Gamma^2_Se^{-{{\Gamma T}\over 4}}{\Re} e^{i\phi^m_B}
\nonumber\\
&& 
\times
\sum_{\alpha=1}^2 \pm |\eta_\alpha|
e^{ \mp ( {{\Delta\Gamma}\over 2} + i\Delta m) {t\over 2}} 
\left[
e^{-(\Gamma_S  - i\Delta m){T\over 2} - i\phi_\alpha }
- |\eta_1\eta_2| e^{-(\Gamma_L  + i\Delta m){T\over 2} 
                       + i\phi_{\bar \alpha}}\right],
\eea
with the upper (lower) sign corresponding to $\alpha=1 (2)$,
and $\bar \alpha =1$ for $\alpha =2$ and vice versa.
The phases $\phi^m_B$, $m=0,1$,
arise from the elements 
$\rho^{PS}_{m0} =|\rho_{m0}^{PS}|e^{i\phi^m_B}$
of the spin density matrix.
For convenience,
we have also introduced the definitions 
$\Delta \Gamma=\Gamma_S-\Gamma_L$,
$\Delta m = m_L-m_S$,
$t = t_2 - t_1$, 
$\Gamma = \Gamma_S+\Gamma_L$,
$m=m_L+m_S$,
and $T=t_1+t_2$. 

In the next section,
we construct an asymmetry $A_{12}(\hat q)$ to
extract parameters sensitive to CP and CPT violation.
However,
we can already note here 
that the term in the production rate proportional to $|M_{SS}|$ 
is independent of $t$
and consequently is absent from the numerator of any asymmetry
defined as a $t$-sensitive difference of production rates.
In contrast,
this term does contribute to the denominator of such an asymmetry,
thereby making the modulation of the signal harder to see.
An incoherent even-parity background merely renormalizes this term, 
further increasing the background 
but not directly influencing the signal.

The decay rate exhibits some interesting and potentially
useful properties.
Consider first the case where $f_1 = f_2$,
i.e., both kaons decay to the same final state.
Then, 
it follows from the structure of the initial wave function
that all terms either independent of or quadratic in the background
(the first and third terms in Eq.\ \rf{ang})
are invariant under the transformation
$t \to - t$,
whereas the linear term
(proportional to $M_{PS}$ in Eq.\ \rf{ang}) 
changes sign.
Among the consequences is that the asymmetry $A_{12}(\hat q)$ 
defined in the next section
contains a term \it linear \rm in the background.
This may provide a means of measuring $\rho^{PS}$,
as discussed below.
If instead we consider the case of different decays 
occurring at the same time $t_1 = t_2$,
then the terms either independent of or quadratic in 
the $S$-wave amplitude are invariant under the transformation
$f_1 \leftrightarrow f_2$,
whereas the linear term changes sign.
Also, as expected, $R(T/2,T/2,\hat q) \propto |M_{SS}|$. 

\vglue 0.4cm
\noindent 
{\it 3.\ Background enhancement.}
The parameters sensitive to CP or CPT violation are best extracted from 
singly and doubly integrated asymmetries.
Define
\beq
R(|t|,\hat q) = R^{+}(|t|,\hat q) - R^{-}(|t|,\hat q)
\eeq
and 
\beq
R(\hat q) = \int_0^{T_L} d|t| R(|t|,\hat q),
\eeq
where 
\beq
R^{\pm}(|t|,\hat q) 
= \int_0^{T_L} dt_1 \int_0^{T_L} dt_2 \delta( t_1 - t_2 \pm |t|)
R(t_1,t_2,\hat q).
\eeq
Here,
the time $T_L$ is related to the acceptance for detecting $K_L$. 
Assuming a maximum distance of $d_L = 5$ m between the
primary vertex and the detector,
together with a boost factor $\ga_K \sim 5$
corresponding approximately to $E_\ga = 10$ GeV photons,
we get
\beq
T_L \Gamma_L = {{d_L}\over \beta_K \gamma_K} \Gamma_L \sim 6.5\times
10^{-2},
\eeq
which leads to $S_L = 1 - \exp(-T_L\Gamma_L) \sim 6\%$ 
acceptance for $K_L$ detection. 

The signal can be parametrized by an asymmetry $A_{12}(\hat q)$, 
defined as 
\beq
A_{12}(\hat q) = {{ \int_0^{T_L} d|t| [R^{+}(|t|,\hat q) - R^{-}(|t|,\hat
q)]} \over 
{\int d^2{\hat q} \int_{0}^{T_L} d|t| [ R^{+}(|t|,\hat q) + R^{-}(|t|,\hat
q)] }}.
\eeq
The detailed form of this asymmetry is somewhat involved. 
For simplicity,
we adopt the approximations 
$\Gamma_S \simeq \Gamma \simeq 2\Delta m \gg \Gamma_L$,
which are well justified experimentally.
With these approximations,
the coefficients in the expressions that follow hold to about 5\%.
Since for either semileptonic or $2\pi$ decays
the ratio $r\equiv\eta_2/\eta_1$ is one to about 0.1\%,
it is also useful to write
$r = 1 + \ka$,
where $\ka \ll 1$ and $\ka$ can be zero for certain final states.
Combining these approximations and definitions,
we find
\beq
A_{12}(\hat q) = { 
{ \left[{{\Gamma_S}\over {\Gamma_L}}\kappa S_L +
2\sin\Delta\phi\right]W_{PP}(\hat q)
+ {1\over 5}\sum_{m=0}^1 \left({{A^m_1}\over {|\eta_1|}} + 
{{A^m_2}\over {|\eta_2|}}\right) W^m_{PS}(\hat q) }
\over 
{\int d^2\hat q\left[  \left( {{\Gamma_S}\over {\Gamma_L}} S_L -
2\cos\Delta\phi\right)W_{PP}(\hat q) 
+ {1\over {2|\eta_1\eta_2|}}W_{SS}(\hat q)\right]}},
\label{p1}
\eeq
where $\Delta\phi=\phi_1-\phi_2$ 
and 
\beq
A^m_i = \cos (\ph^m_\be - \ph_i) + 3 \sin (\ph^m_\be - \ph_i) .
\label{n1}
\eeq

To gain insight into the content of the asymmetry $A_{12}(\hat q)$,
consider first decays into the same final state,
$f_1 = f_2$.
Then,
$r=1$, 
$\ka = 0$,
$\et_1 = \et_2\equiv \eta$, 
and $A^m_1 = A^m_2 \equiv A^m$.
The expression \rf{p1} therefore simplifies considerably.
If the final states consist of the same $2\pi$ mode, 
then we find 
\beq
A_{12}(\hat q) \sim { 
{ { {2} \over {5|\eta|}}\sum_{m=0}^1 A^m  W^m_{PS}(\hat q) }
\over 
{\int d^2\hat q\left[   {{\Gamma_S}\over {\Gamma_L}} S_L 
  W_{PP}(\hat q) + {1\over {2|\eta|^2}}W_{SS}(\hat
    q)\right]}}.
\label{p}
\eeq
This provides a means to measure $\ph$ 
if the background is known
(or possibly a means to determine the background if $\ph$ is known).
For example,
the present bounds on CPT are limited by the precision 
to which $\ph_1$ and $\ph_2$ are known
(roughly, a degree).
Results of similar appearance arise for,
say, 
semileptonic final states.
The situation is also interesting for $f_1 = 2\pi^0$
and $f_2=\pi^+ \pi^-$.
Then, 
$|\et_\al| \simeq 10^{-3}$ and 
$r \simeq 1 + 3 {\Re} \ep^\prime /\ep$.
The issue to resolve is under what circumstances
$W_{SP} \neq 0 $ enhances or masks the signal. 

The existing data on $S$-$P$ interferometry 
with $E_\gamma \lsim 10$ GeV come from two experiments, 
at DESY 
\cite{fr} 
and Daresbury 
\cite{bar}. 
Both experiments clearly identify 
a significant $S$-wave background under the dominant $P$-wave signal. 
Since in diffraction 
$\rho^{11}_{11}$ is the largest element of
the spin density matrix,
one would expect $S$-$P$ interference 
to be dominated by the $W^1_{PS}$ term,
i.e., by $\rho^{10}_{10}$. 
However,
in the limit $t'=t-t_{min}\to 0$,
the $S$-wave production 
is dominated by nucleon helicity-flip 
and so $\rho^{10}_{00}$ should dominate. 
This is consistent with the DESY results where
$t_{max}\sim 0.2$ GeV. 
Nonetheless,
the higher-$t$ data from Daresbury 
($t_{max} =1.6$ GeV) 
do indicate the presence of $\rho^{10}_{10}$,
and this despite
the neglect of its contribution in the partial-wave analysis 
appropriate to the low-$t$ behavior.
Using Eq.\ \rf{p},
it follows that a nonvanishing azimuthal dependence 
in the $S$-$P$ interference 
due to $\rho^{10}_{10}$ would significantly enhance the sensitivity
to the weak phases $\phi_i$. 

High-statistics experiments measuring 
all elements of the spin density matrix are clearly needed. 
An experiment of this type has
recently been proposed for Jefferson Lab 
\cite{ad}.
With the planned accelerator-energy upgrade,
this experiment would have access to approximately 
$10^5$ $\gamma p\to K\bar K p$ events per day 
(compare,
for example, 
the total of $3500$ events collected at DESY). 
Existing experimental results 
fail to provide definitive estimates for the two most 
relevant elements of the spin density matrix,
$\rho^{10}_{00}$ and $\rho^{10}_{10}$.  
The magnitude of $\rho^{10}_{00}$ has 
been estimated to be of the order of a few percent,
while no unique prediction for the phases has been found. 

The scales partially controlling the rates and the asymmetry 
$A_{12}(\hat q)$ are given by
\bea
|W_{PP}| &\sim& O(\rho^{11}_{11}) = 1,
\nonumber\\
|W_{PS}|/|\eta| &\sim& O(|\rho^{10}|)/|\eta| \sim |W_{SS}|/|\eta| \sim 
O(|\rho^{00})/|\eta| \equiv s.
\eea
Examining the expressions given in the above expressions 
for the numerator and denominator of $A_{12}(\hat q)$,
one discovers also a natural scale for $2\pi$ decays
given by
\beq
a = O(\kappa S_L\Gamma_S/\Gamma_L) \sim O(2\sin\Delta\phi) \sim   
O(S_L|\eta|\Gamma_S/\Gamma_L) 
\sim  10^{-2}-10^{-1}.
\eeq
The relative sizes of $s$ and $a$ therefore provide a 
separation into distinct regimes,
to be considered in turn.

If $s < a$,
the asymmetry is dominated by the term 
$ 2\sin\Delta\phi \sim 6 {\Im} {\ep^\prime}/\ep$, 
while sensitivity to the usual 
$3\Re \ep^\prime/\ep$ 
is suppressed due to the acceptance factor $S_L$. 
Terms proportional to the $S$-wave background are 
roughly an order of magnitude smaller.
Thus,
depending on the phase of the coherent background,
it may be feasible to extract information about $\ph_\al$. 

If $a < s$  
the contribution from $s$ dominates the numerator.
As $s$ grows,
the denominator acquires nontrivial $s$ dependence.
This case is potentially very important because
${\ep^\prime}/\ep$
is no longer a factor
and so the whole measurement can focus on
improving the bounds on $\ph_\al$.
If an experiment can be set up in this regime,
it would represent a novel means of measuring
two of the more elusive quantities in CP physics
and of bounding CPT.
This appears to be the most favorable case.
Since $s =  O(|\rho^{10}|)/\vert \et \vert 
\simeq 10^{3} \vert \rho^{10} \vert$,
this corresponds to a $|\rho^{10}|$ magnitude of 
greater than $O(10^{-3})$. 
However, 
it cannot be arbitrarily large, 
since for $s >> a_1 $  
the signal becomes background dominated.
Although there is still information 
in the modulations via the asymmetry,
the signal falls off as $1/s$ due to the 
quadratic contributions in the denominator,
and so measurements become harder.
The current experimental situation seems to favor values of $s$ 
of the order of $O(10^{-2})$,
i.e.,
with the $K_SK_S$ decays from
the $S$ wave dominating the denominator. 

\vglue 0.4cm
\noindent 
{\it 4.\ Simulation.}
In a real experiment,
the limit of the twice-integrated asymmetry would not be taken,
and instead
the complexities of the full system would need to be simulated.
As an initial contribution along these lines,
we have implemented a preliminary Monte-Carlo study 
assuming a flux of $5\times 10^8\gamma/s$ as expected
in the later phase of the proposed photon experiments at Jefferson Lab. 
This translates into $O(10^{10})$ $K_SK_L$ pairs 
from $\phi$ decay per year. 

We assume for definiteness
that the only nonvanishing amplitudes are
the $P$-wave helicity nonflip at both photon and nucleon vertices,
$a^n_1$, 
the $P$-wave single helicity flip at photon and nucleon
vertices, 
$a^f_0$, 
the $S$-wave nucleon helicity nonflip, 
$b^n$,
and $S$-wave single nucleon flip,  
$b^f$. 
In terms of these amplitudes,
the nonvanishing elements of the spin density matrix 
are then given by
\bea
\rho^{11}_{11}&=&\rho^{11}_{-1-1}=2|a^n_1|^2,
\quad 
\rho^{11}_{00}=2|a^f|^2,
\nonumber\\
\rho^{10}_{10}&=&-\rho^{10}_{-10}=2a^n_1b^{f*},
\quad
\rho^{10}_{00}=2a^fb^{f*},
\nonumber\\
\rho^{00}_{00} &=& 2|b^f|^2 + 4|b^n|^2. 
\eea
We adopt the choice $|a^f| \sim |b^f| \sim |b^n| = 0.1$.
Together with $\rho^{11}_{11} \sim 0.5$,
which is fixed by the condition
$1=$ tr$\rho = 2\rho^{11}_{11} + \rho^{11}_{00} + \rho^{00}_{00}$,
this leads to 
$\rho^{11}_{00} \sim \rho^{00}_{00} \sim 0.01$,
which is consistent with the low-$t$ DESY results. 

For simplicity,
we limit attention to the case where both neutral kaons 
decay to $\pi^+\pi^-$ states. 
With the rates given above, 
the $P$ wave contributes 
about $10^4$ $\gamma p \to (K^0\bar K^0)^P p \to 2(\pi^+\pi^-) p$ 
events.
Similarly,
the $(K_SK_S)^S$ in the $S$ wave 
yields approximately $O(10^7)$ events.
We adopt the former as the number of generated events.

\begin{figure}
\centerline{\psfig{figure=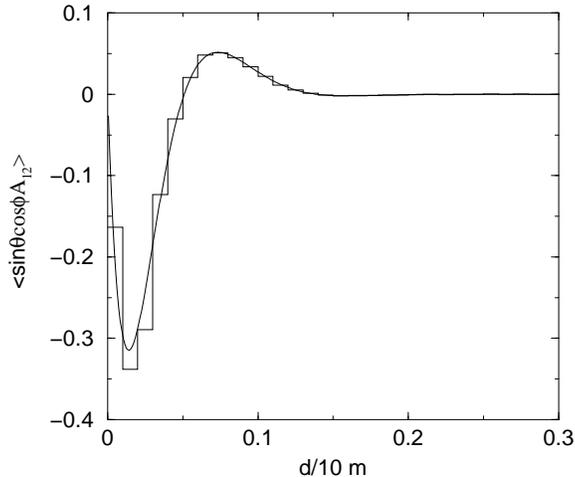,width=0.5\hsize}}
\smallskip
\caption{Theoretical prediction (solid) vs Monte Carlo (steps)
simulations of $A_{12}$ as a function of the relative decay distance
$d$ as measured in the lab.}
\label{fig1}
\end{figure}

\begin{figure}
\centerline{\psfig{figure=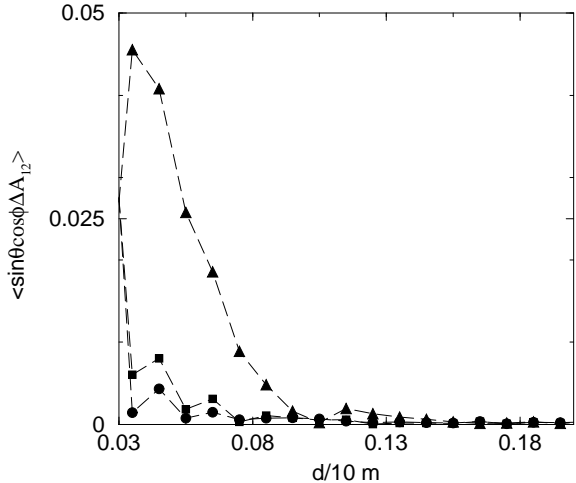,width=0.5\hsize}}
\smallskip
\caption{Sensitivity of the simulated asymmetry to the weak
phase $\phi_\alpha$.}
\label{fig2}
\end{figure}

Under the above conditions,
we have simulated $R(|t|,{\hat q})$ 
and extracted the asymmetry $A_{12}$. 
The result is shown in Fig.\ 1.
The solid and histogram lines correspond
to the theoretical prediction and the simulation,
respectively. 
The sensitivity to the phase $\phi_\alpha=\phi_{+-}$
can be displayed by comparing the magnitudes 
of differences in the asymmetry 
simulated with $\phi^{MC}_\alpha=45^\circ$ 
with theoretical predictions.
In Fig.\ 2,
these differences are plotted for  
three theoretical predictions,
calculated using
$\phi_\alpha = \phi^{MC}$ (circles),
$\phi_\alpha=\phi^{MC} + 1^\circ$ (squares),
and 
$\phi_\alpha = \phi^{MC} + 10^\circ$ (triangles). 
Inspection of the two figures, 
in particular Fig.\ 2, 
suggests that with the anticipated number of events 
it should be possible in principle to extract the weak phases 
to within $O(1^\circ)$ accuracy. 
With a full partial wave analysis,
the sensitivity might be enhanced by another order of magnitude. 

Within the context of conventional quantum field theory,
with CPT violation at the level of the standard model
described by the Lorentz-violating standard-model extension,
the effect of CPT violation on an oscillating neutral meson
depends on the meson velocity magnitude and orientation
\cite{ak}.
However,
in simulating the double-pion decays of the kaons,
we have taken the CPT-sensitive phases to be
independent of orientation for simplicity.
This corresponds to the case where nonzero CPT-violating phases 
are determined by the timelike component of the parameter $\De a_\mu$
in the standard-model extension. 
For this situation,
boosting the meson with a boost factor $\ga$ 
enhances the CPT-violation effect,
inducing a corresponding additive change 
in $\phi_\alpha$ by an amount $\gamma-1$,
i.e., 
$\phi_\alpha(\gamma) = \gamma\phi_\alpha(1)$. 
Thus,
an $O(1^\circ)$ sensitivity to $\phi_\alpha$ 
in a photoproduction experiment with boost factor $\gamma\sim O(10)$ 
is comparable to a $O(0.1^\circ)$ sensitivity to 
$\phi_\alpha$ in a similar experiment at rest. 

It would evidently be interesting to study also
the prospects for enhanced CPT reach
in the more general case of orientation-dependent effects.
The orientation dependence leads to additional possibilities
for CPT signals,
including notably sidereal-time dependence
\cite{ak}.
This has recently been used by the KTeV Collaboration
to obtain a CPT bound in the kaon system
that is independent of all previous bounds
\cite{k99}. 
In the present context,
a complete analysis is likely to require more detailed input 
regarding the angular distribution of the kaon momentum spectrum
and the detector performance. 
With a favorable scenario, 
bounds on all four independent coefficients for CPT violation
in the kaon system could be obtained.

\vglue 0.4cm
\noindent 
{\it 5.\ Summary.}
This work has investigated some aspects of the CP and CPT sensitivity
that could be attained by experiments
involving photoproduction of neutral-kaon pairs.
The photoproduction mechanism generates a coherent $P$ wave as usual,
but also yields a coherent $S$ wave along with an incoherent background.
The magnitude of the coherent $S$ wave
is presently uncertain but could be significant.
Under favorable circumstances,
the resulting CPT sensitivity could be comparable 
to that attainable at conventional $\ph$ factories.
Both analytical calculations and Monte-Carlo simulation
indicate that interference between the $P$ and $S$ waves 
might lead to an enhancement of an order of magnitude 
in the existing CPT reach.

\vglue 0.4cm
We thank A.\ Dzierba for discussions.
This work was supported in part
by the United States Department of Energy 
through grants DE-FG02-91ER40661, DE-FG02-87ER40365
and contract DE-AC05-84ER40150,
under which the Southeastern Universities Research Association
(SURA) operates the Thomas Jefferson National Laboratory
(Jefferson Lab).

\vglue 0.4cm

\end{document}